%% file: root.tex
\documentclass{ifacconf}
\usepackage{natbib}        
\usepackage{float}
\usepackage{graphicx}
\usepackage{amsfonts,amsmath}
\usepackage{multirow}
\usepackage{mathtools}
\usepackage[most]{tcolorbox}
\usepackage{xcolor}
\usepackage{adjustbox}
\usepackage{algorithm}
\usepackage{enumitem}
\usepackage{matlab-prettifier}
\usepackage{courier}
\usepackage{url}
\usepackage{siunitx}
\usepackage{tabularx}

\newcommand{\nx}{n_\mathrm{x}}
\newcommand{\dnu}{n_\mathrm{u}}
\newcommand{\ny}{n_\mathrm{y}}
\newcommand{\np}{n_\mathrm{p}}
\newcommand{\diff}{\xi}
\newcommand{\R}{\mathbb{R}}
\newcommand{\Pset}{\mathcal{P}}
\newcommand*\crule[3]{%
  {\color[rgb]{#1}\rule[0.4ex]{#2}{#3}}}

\DeclarePairedDelimiterX{\norm}[1]{\lVert}{\rVert}{#1}

\begin{document}
\begin{frontmatter}

    \title{Automated Linear Parameter-Varying Modeling of Nonlinear Systems: A Global Embedding Approach\thanksref{footnoteinfo}}

    \thanks[footnoteinfo]{This research was supported by the European Space Agency (grant number: 4000145530) and The MathWorks Inc. Opinions, findings, conclusions or recommendations expressed in this abstract are those of the authors and do not necessarily reflect the views of The MathWorks Inc. or the European Space Agency.}

    \author[TUE]{E. Javier Olucha}
    \author[TUE]{Patrick J. W. Koelewijn}
    \author[TUE]{Amritam Das}
    \author[TUE, HUNREN]{Roland Tóth}

    \address[TUE]{Control Systems Group, Eindhoven University of Technology, Eindhoven, The Netherlands \\(e-mail: \{e.j.olucha.delgado, am.das, r.toth\}@tue.nl).}
    \address[ HUNREN]{Systems and Control Laboratory, HUN-REN Institute for Computer Science and Control, Budapest, Hungary}

    \begin{abstract}                
        In this paper, an automated \emph{linear parameter-varying} (LPV) model conversion approach is proposed for nonlinear dynamical systems. The proposed method achieves global embedding of the original nonlinear behavior of the system by leveraging the second fundamental theorem of calculus to factorize matrix function expressions without any approximation. The implementation of the proposed method in the \textsc{LPVcore} toolbox for \textsc{Matlab} is discussed, 
        and its performance is showcased on a comprehensive example of automated LPV model conversion of an unbalanced disk system, which is then used to design an LPV controller that is deployed on the original nonlinear system. In addition, the conversion capabilities are further demonstrated by obtaining an LPV embedding of a three-degree-of-freedom control moment gyroscope. All software implementations are available at \url{www.lpvcore.net}.
    \end{abstract}


\end{frontmatter}

\input{chapters/intro}

\input{chapters/problem_def}
\input{chapters/method}
\input{chapters/implementation.tex}

\input{chapters/examples}

\input{chapters/conclusion}

\bibliography{LPVS25,ieeetr}
\end{document}

%% file: chapters/intro.tex
\section{Introduction\label{sec:intro}}

The general concept of \emph{linear parameter-varying} (LPV) systems has been introduced to provide
an easily applicable and deployable analysis and control synthesis framework for \emph{nonlinear} (NL) systems based on the extension of powerful approaches of the \emph{linear time-invariant} (LTI) framework \citep{mohammadpour2011lpv,Toth2010SpringerBook,APKARIAN19951251}. In LPV systems, the signal relations between inputs and outputs are considered linear but, at the same time, dependent on a so-called \emph{scheduling variable} $p$. 
In this way, the variation of $p$ represents changing operating conditions, nonlinear effects, etc., described by an underlying \emph{scheduling map}, and aims at complete embedding of the original NL behavior into the solution set of an LPV system representation \citep{Toth2010SpringerBook,Rugh00}. While the former objective is pursued by the so-called \emph{global} LPV modeling approaches; alternatively, one can aim for only an approximation of the NL behavior by interpolation of various linearizations of the system around operating points or signal trajectories, often referred to as \emph{local} modeling; see, \emph{e.g.}, \citep{Toth14JPC,Shamma90c}.

Although many practical approaches have been introduced for local LPV modeling and have also been implemented in various software packages such as \textsc{Matlab}, the development of global LPV modeling methods has been limited, especially in terms of widely usable software implementations.
Existing approaches for \emph{global} LPV modeling of NL dynamical systems can be classified into two main categories: \emph{substitution based transformation} (SBT) methods \citep{Shamma93, Papageorgiou00,Bokor2007,Toth11ACC_Philips,Rugh00,Leith98b,Marcos04} and \emph{automated conversion procedures} \citep{Donida2009,Kwiatkowski06,Kwiatkowski4494453,Toth2010SpringerBook,Hoffmann2015,Toth19TAC,Toth20bIET,ROTONDO201544}. Specifically, \cite{Donida2009} proposed a promising symbolic conversion method based on \textsc{Modelica}, but the approach has never been released as an available software package. In \cite{Toth19TAC}, a multipath feedback linearization-based LPV model conversion method has been proposed that provides global embedding guarantees, but can generally result in dependence of $p$ on various derivatives of the input and output of the system.
Summand decomposition and factorization of nonlinear terms have been proposed in various forms such as in \cite{Kwiatkowski06,Toth2010SpringerBook,ROTONDO201544}, resulting in applicable conversion methods, which also allow choosing between complexity and conservativeness of possible LPV embeddings. However, these techniques need the exploration of all possible decomposition and division possibilities, leading to decision trees with a combinatorial explosion of possibilities and suffering from high computational load due to the involved symbolic manipulations and lack of reliable software implementation. In \cite{Hoffmann2015}, a \emph{linear matrix inequalities}-based optimization approach has been proposed to efficiently solve the choice of optimal trade-off between the combination of various conversion possibilities. Even if this provided a powerful tool, it is limited to small or moderate state and scheduling dimensions and requires pre-proposed conversion terms. All these problems have been overcome by data-driven techniques \citep{Kwiatkowski4494453,Toth20bIET}, representing an evolution of conversion methods based on \emph{principal components analysis} (PCA), capable of automatic determination of the relevant scheduling dimension and the construction of a data-based scheduling map. These methods have also been implemented in toolboxes such as \textsc{LPVcore}. However, success of these modeling techniques depends largely on the available data sets and they are inherently approximative in their nature. Hence, in general there is a lack of reliable and efficiently computable analytic LPV model conversion methods that ensure global embedding of the original NL system and available in terms of an off-the-shelf software solution.

Recently, the idea of a novel LPV modeling procedure has been proposed in the PhD thesis \citep{koelewijnAnalysisControlNonlinear2023} for general NL models using the \emph{Second Fundamental Theorem of Calculus} (FTC) to factorize matrix function expressions without any approximation and alleviating the need for summand decompositions, decision trees, or other complex embedding processes. Then, the approach has been successfully applied to construct a solution to various problems, such as in \citep{Toth23CDCj}, but has never been properly formulated and discussed as a global LPV embedding and model conversion tool and no software implementation has been developed for this purpose. In this paper, our main contributions are as follows:
\begin{itemize}
    \item Mathematically rigorous formulation of the global LPV embedding of a rather general class of NL systems using an FTC-based factorization of nonlinearites.
    \item Analysis of the properties of the resulting systematic LPV modeling process, discussing its pros and cons.
    \item  Developing a publicly available software implementation of the proposed FTC-based conversion in the toolbox \textsc{LPVcore} and demonstrating the capabilities of the approach in practical examples.
\end{itemize}

The paper is structured as follows: In Section~\ref{sec:probdef}, the problem of global embedding-based LPV model construction for nonlinear systems is introduced. For this problem, the proposed concept of FTC-based model conversion is proposed in Section~\ref{sec:method}, also discussing its properties. Then, in Section~\ref{sec:implementation} implementation of the approach in \textsc{Matlab} is discussed, followed in Section~\ref{sec:examples} by the demonstration of the conversion capabilities of the developed tooling on LPV model-based controller design for an unbalanced disk system and LPV modeling of a three-degree-of-freedom control moment gyroscope. Finally, in Section~\ref{sec:conclusion}, the main conclusions on the achieved results are drawn.

%% file: chapters/problem_def.tex
\section{Problem definition\label{sec:probdef}}

Consider a nonlinear dynamical system given by the \emph{state-space} (SS) representation 
\begin{subequations}
	\label{eq:nl_dyn}
\begin{align}
		\diff x(t) & = f(x(t), u(t)), \\
		y(t)       & = h(x(t), u(t)),
\end{align}
\end{subequations}
where $t \in \mathbb{T}$ is time, $\diff$ is $\diff x(t)= \frac{d}{dt}x(t)$ in the continuous-time case with $\mathbb{T}=\mathbb{R}$ and $\diff x(t)= x(t+1)$ in the discrete-time case with $\mathbb{T}=\mathbb{Z}$, $x(t) \in \mathcal{X} \subseteq \R^{\nx}$, $u(t) \in \mathcal{U} \subseteq \R^{\dnu}$ and $y(t) \in \mathcal{Y} \subseteq \R^{\ny}$ with $\nx, \dnu, \ny \in \mathbb{N}$ are the state, input, and output signals associated with the system, respectively. The functions $f: \R^{\nx} \times \R^{\dnu} \to \R^{\nx}$ and $h: \R^{\nx} \to \R^{\ny}$ are continuously differentiable once, i.e., $f, h \in \mathcal{C}_1$, and the solutions of \eqref{eq:nl_dyn} have left compact support, they are forward complete and unique, i.e., for any initial condition $x(t_0)$ and any input trajectory $u(t)$, the solutions of \eqref{eq:nl_dyn} are uniquely determined for all $t \geq t_0$. Moreover, $(0, 0) \in \mathcal{X} \times \mathcal{U}$. Finally we denote the behavior of \eqref{eq:nl_dyn}, i.e., the set of all possible trajectories by
\begin{equation}
	\label{eq:nl_behavior}
	\mathcal{B} \coloneq \left\{ (x, u, y) \in \left(\mathcal{X}, \mathcal{U}, \mathcal{Y}\right)^{\mathbb{T}} \mid (x, u, y) ~ \text{satisfy}~\eqref{eq:nl_dyn} \right\},
\end{equation}
where $\mathcal{X}^{\mathbb{T}}$ denotes the set of all signals $\mathbb{T} \rightarrow \mathcal{X}$ with left compact support.

In this paper, our objective is to automatically convert the system description \eqref{eq:nl_dyn} into a \emph{linear parameter-varying} (LPV) representation of the form
\begin{subequations}
	\label{eq:lpv_dyn}
	\begin{align}
		\diff x(t) & = A(p(t))x(t) + B(p(t))u(t); \\
		y(t)       & = C(p(t))x(t) + D(p(t))u(t),
	\end{align}
\end{subequations}
where $p \in \Pset \subseteq \R^{\np}$ with $\np \in \mathbb{N}$ is the scheduling variable, $A : \Pset \rightarrow \R^{\nx \times \nx}$, $B : \Pset \rightarrow \R^{\nx \times \dnu}$, $C : \Pset \rightarrow \R^{\ny \times \nx}$, and $D : \Pset \rightarrow \R^{\ny \times \dnu}$ are smooth real-valued matrix functions, and the solutions of \eqref{eq:lpv_dyn} with left compact support are forward complete and unique. For a given scheduling trajectory $p(t)\in\mathcal{P}$, the behavior of \eqref{eq:lpv_dyn} is defined as
\begin{equation}
	\label{eq:p_behavior}
	\mathcal{B}_p \coloneq \left\{ (x, u, y) \in \left(\mathcal{X}, \mathcal{U}, \mathcal{Y}\right)^{\mathbb{T}} \mid (x, u, y, p) ~ \text{satisfy}~\eqref{eq:lpv_dyn} \right\},
\end{equation}
and the behavior of \eqref{eq:lpv_dyn} for all possible scheduling trajectories is defined as
\begin{equation}
	\label{eq:lpv_behavior}
	\mathcal{B}_{\text{LPV}} \coloneq \bigcup_{p \in \Pset^{\mathbb{T}}} \mathcal{B}_p(p).
\end{equation}

We consider the LPV representation \eqref{eq:lpv_dyn} to be a so-called global LPV embedding of \eqref{eq:nl_dyn} if in addition, we can construct a so-called \emph{scheduling map} $\eta : \mathcal{X} \times \mathcal{U} \rightarrow \Pset$,
\begin{equation}
	\label{eq:scheduling_map}
	p(t) = \eta(x(t), u(t)),
\end{equation}
such that
\begin{subequations}
\label{eq:LPV_realization}
	\begin{align}
		f(x, u) & = A(\eta(x, u)) x + B(\eta(x, u)) u, \\
		h(x, u) & = C(\eta(x, u)) x + D(\eta(x, u)) u, 
	\end{align}
\end{subequations}
for all $(x, u) \in \mathcal{X} \times \mathcal{U}$. This gives that $\eta(\mathcal{X},\mathcal{U})\subseteq \mathcal{P} \subseteq \mathbb{R}^{n_\mathrm{p}}$, where $\mathcal{P}$ is often chosen be a compact convex set, if the model is further utilized for analysis or control synthesis.
Consequently, the behavior of the nonlinear system is included (embedded) in the behavior of the LPV system, i.e., $\mathcal{B} \subseteq \mathcal{B}_{\text{LPV}}$. In the next section, we will discuss the proposed method to achieve this objective.


%% file: chapters/method.tex
\section{Method\label{sec:method}}
To achieve our objective of automatically converting the nonlinear system description \eqref{eq:nl_dyn} into a global LPV embedding given by \eqref{eq:lpv_dyn} with $p(t) = \eta(x(t), u(t))$, we leverage the \emph{Second Fundamental Theorem of Calculus} (FTC)~\citep{apostolCalculus1Onevariable1980} for function factorization as proposed in~\citep[Appendix~C.1]{koelewijnAnalysisControlNonlinear2023}.

Consider a function $g(z) = [g_1(z) \ \cdots \ g_m(z)]^\top$ where $g_i(z) : \R^{n} \rightarrow \R^{m}$ for $i = 1,\dots,m$ and $z \in \R^{n}$. For $z, \, z_{\ast} \in \R^\mathrm{n}$, we define the auxiliary function
\begin{equation}
	\label{eq:auxiliary}
	\bar{g}_i(\lambda) = g_i(z_{\ast} + \lambda(z - z_{\ast})),
\end{equation}
for $i = 1,\dots,m$ and $\lambda \in [0, 1]$. Then by the FTC, we have that
\begin{equation}
	\label{eq:FTC}
	\bar{g}_i(1) - \bar{g}_i(0) = \int_{0}^{1} \frac{d \bar{g}_i}{d \lambda}(\lambda) \, d \lambda,
\end{equation}
and using the relations $\bar{g}_i(1)=g_i(z)$ and $\bar{g}_i(0) =  g_i(z_{\ast})$, obtained by evaluating \eqref{eq:auxiliary} at $\lambda = \{0, 1\}$, it follows that
\begin{equation}
	\label{eq:FTC2}
	g_i(z) - g_i(z_{\ast}) = \int_{0}^{1} \frac{d \bar{g}_i}{d \lambda}(\lambda) \, d \lambda.
\end{equation}
Now, by taking the chain rule of the integrand in \eqref{eq:FTC2}:
\begin{equation}
	\label{eq:chain_rule}
	\frac{d \bar{g}_i}{d \lambda} = \frac{\partial g_i}{\partial z} \frac{d \bar{g}_i}{d \lambda} = \frac{\partial g_i}{\partial z}(z_{\ast} + \lambda(z - z_{\ast}))(z - z_{\ast}),
\end{equation}
where $\frac{\partial g_i}{\partial z}$ is the transpose of the gradient of $g_i$, and by substituting \eqref{eq:chain_rule} into \eqref{eq:FTC2}, we end up with
\begin{equation}
	g_i(z) - g_i(z_{\ast}) = \int_{0}^{1} \frac{\partial g_i}{\partial z}(z_{\ast} + \lambda(z - z_{\ast})) \, d \lambda \, (z - z_{\ast}),
\end{equation}
for $i = 1,\dots,m$. Then, combining the elements, $g$ is expressed as
\begin{align}
	\label{eq:fun_factorization}
	g(z) - g(z_{\ast}) & = \begin{bmatrix}
		                      g_1(z) - g_1(z_{\ast}) \\
		                      \vdots                \\
		                      g_m(z) - g_m(z_{\ast})
	                      \end{bmatrix}; \notag                                                                                         \\
	                  & = \begin{bmatrix}
		                      \int_{0}^{1} \frac{\partial g_1}{\partial z}(z_{\ast} + \lambda(z - z_{\ast})) \, d \lambda \, (z - z_{\ast}) \\
		                      \vdots                                                                                       \\
		                      \int_{0}^{1} \frac{\partial g_m}{\partial z}(z_{\ast} + \lambda(z - z_{\ast})) \, d \lambda \, (z - z_{\ast})
	                      \end{bmatrix}; \notag \\
	                  & = \int_{0}^{1} \frac{\partial g}{\partial z}(z_{\ast} + \lambda(z - z_{\ast})) \, d \lambda \, (z - z_{\ast}),
\end{align}
where $\frac{\partial g}{\partial z}(z_{\ast} + \lambda(z - z_{\ast})) \in \R^{m \times n}$ is the Jacobian of $g$ evaluated at $z_{\ast} + \lambda(z - z_{\ast})$ and the integral of the Jacobian is taken element-wise.

Now, we can apply \eqref{eq:fun_factorization} to factorize the functions $f$ and $h$ in \eqref{eq:nl_dyn} as follows:
\begin{subequations}
	\label{eq:factorized_nl}
	\begin{align}
		f(x, u) - f(\bar{x}, \bar{u}) & = \bar{A}(x, u) (x - \bar{x}) + \bar{B}(x, u) (u - \bar{u}); \\
		h(x, u) - h(\bar{x}, \bar{u}) & = \bar{C}(x, u) (x - \bar{x}) + \bar{D}(x, u) (u - \bar{u}),
	\end{align}
	where
	\begin{align*}
		\bar{A}(x, u) & = \int_{0}^{1} \frac{\partial f}{\partial x}\left(\bar{x} + \lambda (x - \bar{x}), \bar{u} + \lambda (u - \bar{u})\right) \, d \lambda;  \\
		\bar{B}(x, u) & = \int_{0}^{1} \frac{\partial f}{\partial u}\left(\bar{x} + \lambda (x - \bar{x}) , \bar{u} + \lambda (u - \bar{u})\right) \, d \lambda; \\
		\bar{C}(x, u) & = \int_{0}^{1} \frac{\partial h}{\partial x}\left(\bar{x} + \lambda (x - \bar{x}), \bar{u} + \lambda (u - \bar{u})\right) \, d \lambda;  \\
		\bar{D}(x, u) & = \int_{0}^{1} \frac{\partial h}{\partial u}\left(\bar{x} + \lambda (x - \bar{x}), \bar{u} + \lambda (u - \bar{u})\right) \, d \lambda.
	\end{align*}
\end{subequations}

Moreover, it is often the case that $f(0, 0) = 0$ and {$h(0, 0) = 0$}, i.e., that the origin is an equilibrium point of the system. Otherwise, we can perform a coordinate transformation to ensure this property. In these cases, we can further simplify \eqref{eq:factorized_nl} by taking $\bar{x} = 0$ and $\bar{u} = 0$, resulting in:
\begin{subequations}
	\label{eq:simplified_factorized_nl}
	\begin{align}
		f(x, u) & = \bar{A}(x, u) x + \bar{B}(x, u) u, \\
		h(x, u) & = \bar{C}(x, u) x + \bar{D}(x, u) u,
	\end{align}
	where
	\begin{align*}
		\bar{A}(x, u) \hspace{-1pt} & = \hspace{-2pt} \int_{0}^{1} \! \frac{\partial f}{\partial x}(\lambda x, \lambda u) \, d \lambda, \
		\bar{B}(x, u)  = \hspace{-2pt} \int_{0}^{1} \! \frac{\partial f}{\partial u}(\lambda x, \lambda u) \, d \lambda,                  \\
		\bar{C}(x, u) \hspace{-1pt} & = \hspace{-2pt} \int_{0}^{1} \! \frac{\partial h}{\partial x}(\lambda x, \lambda u) \, d \lambda, \
		\bar{D}(x, u)  = \hspace{-2pt} \int_{0}^{1} \! \frac{\partial h}{\partial u}(\lambda x, \lambda u) \, d \lambda.
	\end{align*}
\end{subequations}
To obtain an LPV form \eqref{eq:lpv_dyn} for a given factorization of $f$ and $h$ in \eqref{eq:simplified_factorized_nl}, a 
scheduling map $\eta$ in terms of \eqref{eq:scheduling_map} is constructed  
%
such that the matrix functions $A,B,C,D$ resulting from \eqref{eq:LPV_realization} via 
\begin{equation}
	\begin{aligned}
		\label{eq:lpv_matrices}
		A(\overbrace{\eta(x,u)}^{p}) & = \bar{A}(x, u), & B(\overbrace{\eta(x,u)}^{p}) & = \bar{B}(x, u), \\
		D(\underbrace{\eta(x,u)}_p) & = \bar{C}(x, u), & D(\underbrace{\eta(x,u)}_p) & = \bar{D}(x, u).
	\end{aligned}
\end{equation}
have a  
%
desired type of scheduling dependency, e.g., affine, polynomial, or rational. An LPV representation with an affine scheduling dependency can for example be obtained by choosing $\eta$ to contain the nonlinear elements of the matrix functions $\bar{A},\, \bar{B},\, \bar{C}$, and $\bar{D}$. 

Through this process, the nonlinear system represented by \eqref{eq:nl_dyn} is directly embedded into a global LPV model of the form \eqref{eq:lpv_dyn} with \eqref{eq:scheduling_map} satisfying \eqref{eq:LPV_realization} via \eqref{eq:simplified_factorized_nl} and \eqref{eq:lpv_matrices} and hence achieving $\mathcal{B} \subseteq \mathcal{B}_{\text{LPV}}$.
An important property of factorization \eqref{eq:simplified_factorized_nl}, compared to other conversion methods, is that there is no approximation involved; hence, embedding of the complete nonlinear behavior $\mathcal{B}$ is guaranteed on the target sets $\mathcal{X}\times\mathcal{U}$. Also, no decisions are required to be made that which functional terms are split or substituted in which order as in other conversion concepts (e.g., \cite{Kwiatkowski06}). Furthermore, the factorization \eqref{eq:simplified_factorized_nl}, i.e., the matrix functions $\bar{A},\ldots,\bar{D}$ are well defined for $x=0$ and $u=0$, because \[\lim_{(x,u)\rightarrow (0,0)} \bar{A}(x,u)=\lim_{\tau \rightarrow 0} \int_{0}^{\tau} \! \frac{\partial f}{\partial x}(\lambda x, \lambda u) \, d \lambda = \frac{\partial f}{\partial x}(0, 0),\]
which exists for all $\bar{A},\ldots,\bar{D}$ as $f,h\in\mathcal{C}_1$. 

In addition to the advantages of the embedding process, a clear price to pay for automatic conversion is that the required information for the scheduling map $\eta$ cannot be controlled during the conversion. For real-world applications, the matrix functions $\bar{A},\ldots,\bar{D}$ often only depend on certain elements of $x$ and $u$, therefore only those elements are required for the scheduling map $\eta$. This will also be exemplified in Section \ref{sec:examples}. On the other hand, mathematically it is possible that the entire $(x,u)$ vector is required to compute the resulting $p$. Furthermore, the resulting dimension of $p$, that is, $n_\mathrm{p}$, depends on the chosen dependency class of $A,\ldots,B$. The more simplistic dependence, e.g., affine, is chosen for these matrix functions, the more complex and higher dimensional the resulting scheduling map $\eta$ becomes. These disadvantages can be efficiently mitigated by recently introduced scheduling reduction techniques, such as \cite{Toth20ACC,Toth16TCST,8431912}, which can often provide dramatic reduction of $n_\mathrm{p}$ without a significant loss of accuracy of the representation; see \cite{olucha2024reductionlinearparametervaryingstatespace} for a comparative study.         

In the next section, we provide the details of the implementation of the proposed method in the \textsc{LPVcore} toolbox.

%% file: chapters/implementation.tex
\section{Implementation in \textsc{LPVcore}\label{sec:implementation}}
The integration of the automatic global LPV embedding method into the \textsc{LPVcore} toolbox is achieved by introducing a set of objects and functions that allow the user to define the nonlinear dynamical system and execute the conversion. As \textsc{Matlab} does not have a native implementation to define systems in the form of \eqref{eq:nl_dyn}, \textsc{LPVcore} introduces the \lstinline{nlss} class. For instance, consider the following nonlinear system
\begin{equation}
    \label{eq:nl_dyn_example}
    \begin{aligned}
        \diff x(t) & = -x(t) + u(t); \\
        y(t)       & = \tanh(x(t)).
    \end{aligned}
\end{equation}
A representation of \eqref{eq:nl_dyn_example} can be created by defining $f$ and $h$ as function handles, providing the system dimensions, and setting the sample time as \mbox{$T_\mathrm{s} = 0$}, \mbox{$T_\mathrm{s} > 0$} or as \mbox{$T_\mathrm{s} = -1$} to indicate that the system is in continuous-time, in discrete-time with the specified sampling time, or in discrete-time with undefined sampling time, respectively:
\begin{lstlisting}[basicstyle=\ttfamily]
f = @(x, u) -x + u;
h = @(x, u) tanh(x);
nx = 1; nu = 1; ny = 1; Ts = -1;
sys = LPVcore.nlss(f, h, nx, nu, ny, Ts);
\end{lstlisting}
%
%
Next, the \textsc{LPVcore} function \lstinline{nlss2lpvss} implements the automatic global LPV embedding procedure described in Section~\ref{sec:method}. This function allows the user to convert \lstinline{nlss} objects into LPV \emph{state-space} (LPV-SS) models, represented by an \textsc{LPVcore} \lstinline{lpvss} object. The LPV-SS models produced by \lstinline{nlss2lpvss} have by default affine dependency on the scheduling variables. This comes from the fact that the function $\eta$ is constructed by including all the nonlinear relationships resulting from~\eqref{eq:simplified_factorized_nl}. 

Continuing with the example, the \lstinline{sys} object that represents \eqref{eq:nl_dyn_example} can be converted into an LPV-SS model by executing
\begin{lstlisting}[linewidth=1.01\columnwidth, basicstyle=\ttfamily]
[lpvsys, eta] = LPVcore.nlss2lpvss(sys,...
    'analytical', 'element')
\end{lstlisting}
The second input argument of \lstinline{nlss2lpvss} determines whether the integration of the Jacobians in \eqref{eq:simplified_factorized_nl} is performed \lstinline{'numerically'} or \mbox{\lstinline{'analytically'}}, as follows:
\begin{itemize}
    \item \textit{Numerical integration (\lstinline{'numerically'} option):} in this mode, each nonlinear element of the Jacobians in \eqref{eq:simplified_factorized_nl} is extracted and associated with an affine scheduling variable. The integrals of each of these elements are used to define the scheduling map $\eta$ and are stored unsolved in the form of the \textsc{Matlab} data type \emph{function handle}, with $(x, u)$ as function input arguments. Then, when evaluating the scheduling map, the \textsc{Matlab} function \lstinline{integral} is invoked to execute the numerical integration using global adaptive quadrature with default error tolerances.
    \item \textit{Analytical integration (\lstinline{'analytically'} option):} in contrast, this mode first integrates the elements of the Jacobians symbolically using the \textsc{Matlab} function \lstinline{int} using the \textsc{Matlab} Symbolic Toolbox. In this case, the third input argument of \lstinline{nlss2lpvss}determines how are the nonlinear elements from the integrated Jacobians extracted. The option \mbox{\lstinline{'element'}} follows the same procedure as the one described for the numerical integration, except the resulting scheduling map contains the analytic solution of the integrals. Alternatively, the option \mbox{\lstinline{'factor'}} uses an algorithm that tries to factor each of the nonlinear elements as a sum of scheduling variables, and then creates the scheduling map based on these scheduling variables.
\end{itemize}
  While an analytical integration produces an exact LPV embedding, it is computationally more expensive than a numerical integration due to the symbolic computations and it can fail to find the antiderivative for some complicated expressions. Lastly, the first output of \lstinline{nlss2lpvss} is the obtained LPV-SS model as an \textsc{LPVcore} \lstinline{lpvss} object, and the second output is the obtained scheduling map $\eta$ as a scheduling map object \lstinline{schedmap}, a class implemented in \textsc{LPVcore}.

\begin{figure}[t]
    \centering
    \includegraphics[width=0.65\columnwidth]{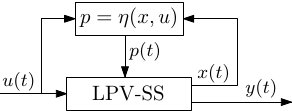}
    \caption{Schematic representation of a self-scheduled simulation with an LPV-SS model and a scheduling map.}
    \label{fig:selfsched_diagram}
\end{figure}
Note that once the LPV-SS model is obtained, the user can readily use it with the remaining functionalities of the \textsc{LPVcore} toolbox. For instance, the \lstinline{lpvss} object can be used for LPV controller design, analysis or model reduction. Moreover, the \textsc{LPVcore} function \lstinline{lsim} has been extended to execute self-scheduled simulations of LPV-SS models, as illustrated in Fig.~\ref{fig:selfsched_diagram}. This can be done by providing a \lstinline{schedmap} object as the second input argument, as follows:
\begin{lstlisting}[basicstyle=\ttfamily]
y = lsim(lpvsys, eta, input, time, ...
    'Solver', @ode45)
\end{lstlisting}
For a complete overview of the available functionalities, consult the documentation which is included in \textsc{LPVcore}.


%% file: chapters/examples.tex
\section{Examples \label{sec:examples}}
In this section, the capabilities of the proposed LPV embedding method are demonstrated through two examples. The first example involves a simple nonlinear model of an unbalanced disk. This example is used to analyze the LPV model obtained with the proposed approach, evaluate the accuracy of the LPV embedding via time-domain simulations, and to design an LPV controller using the obtained model. The second example considers a more complex system to assess the scalability of the proposed method. Specifically, the Quanser 3DOF \emph{control moment gyroscope} (CMG) is considered, and the performance of the method is demonstrated through time-domain simulations.

\subsection{Unbalanced disk \label{sec:unbalanced_disk}}
\begin{figure}[t]
    \centering
    \includegraphics[width=0.4\columnwidth]{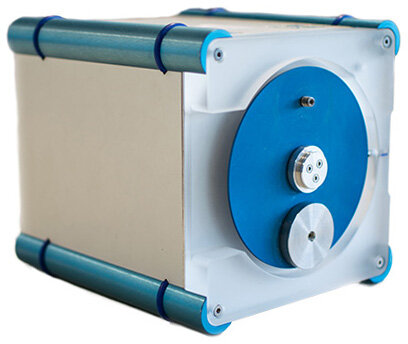}
    \caption{Picture of the unbalanced disk system.}
    \label{fig:unbalanced_disk}
\end{figure}

The unbalanced disk system consists of an actuated disk with one degree of freedom with a mass mounted on it, as shown in Fig.~\ref{fig:unbalanced_disk}. The motion of the unbalanced disk can be described in the form of \eqref{eq:nl_dyn} by
\begin{equation} \label{eq:unbalanced_disk}
    \hspace{-2mm}  \Pi_\mathrm{NL} \coloneq \left\{
    \begin{aligned}
        \dot{x}_1(t) & = x_2(t);                                                                                   \\
        \dot{x}_2(t) & = \tfrac{M g l}{J} \sin(x_1(t)) - \tfrac{1}{\tau} x_2(t) + \tfrac{K_\mathrm{m}}{\tau} u(t); \\
        y(t)         & = x_1(t),
    \end{aligned}\right.
\end{equation}
where $x_1$ and $x_2$ represent the angle of the disk in radians and its angular velocity in radians per second, respectively, $u$ the input voltage to the motor in Volts, $y$ the output of the system, and $M = \num{7e-2} \ \mathrm{kg}$, $g = \num{9.8} \ \mathrm{m \cdot s^{-2}}$, $l = \num{4.2e-2} \ \mathrm{m}$, $J = \num{2.2e-4} \ \mathrm{kg \cdot m^2}$, $\tau = \num{5.971e-1} \ \mathrm{s}$, and $K_\mathrm{m} = \num{1.531e1} \ \mathrm{rad \cdot s^{-1} \cdot V^{-1}}$ are the physical parameters.
%


For illustration, we apply the proposed approach manually first and then we compare the resulting LPV model to the model obtained by the implementation in \textsc{LPVcore}. By applying~\eqref{eq:simplified_factorized_nl} to~\eqref{eq:unbalanced_disk}, we get
\begin{subequations}
    \begin{align}
        \dot{x}(t) & = \bar{A}(x, u)x(t) + \bar{B}(x, u) u(t); \\
        y(t)       & = \bar{C}(x, u) x(t),
    \end{align}
    with
    \begin{align*}
        \bar{A}(x, u) & = \begin{bmatrix}
                              0                                                                          & 1               \\
                              \frac{M g l}{J} \! \int_{0}^{1} \cos\left(\lambda x_1(t) \right) d \lambda & -\frac{1}{\tau}
                          \end{bmatrix};                    \\
        \bar{B}(x, u) & = \begin{bmatrix}
                              0 & \frac{K_\mathrm{m}}{\tau}
                          \end{bmatrix}^\top; \qquad \bar{C}(x, u)                                                       = \begin{bmatrix}
                                                                                                                               1 & 0
                                                                                                                           \end{bmatrix},
    \end{align*}
\end{subequations}
where $x = \operatorname{col}(x_1, x_2)$. The analytical solution of the integral provides $\int_{0}^{1} \cos\left(\lambda x_1(t) \right) \; d \lambda = \frac{\sin(x_1(t))}{x_1(t)} = \operatorname{sinc}(x_1(t))$. Then, we can define the scheduling map in \eqref{eq:scheduling_map} to be $\eta(x, u) = \operatorname{sinc}(x_1)$ to obtain the following  LPV representation with affine coefficient dependence:
\begin{subequations} \label{eq:unbalanced_LPV_manual}
    \begin{equation}
        \hspace{-2mm} \Pi_\mathrm{LPV}^\mathrm{ana} \coloneq \left\{
        \begin{aligned}
            \dot{x}(t) & = \begin{bmatrix}
                               0 & 1 \\ \frac{M g l}{J} p(t) & -\frac{1}{\tau}
                           \end{bmatrix}x(t) + \begin{bmatrix}
                                                   0 \\ \frac{K_\mathrm{m}}{\tau}
                                               \end{bmatrix} u(t); \\
            y(t)       & = \begin{bmatrix}
                               0 & 1
                           \end{bmatrix} x(t),
        \end{aligned} \right.
    \end{equation}
    with
    \begin{align}
        p(t) & = \operatorname{sinc}(x_1(t)),
    \end{align}
\end{subequations}
which is an LPV embedding of~\eqref{eq:unbalanced_disk} with $\mathcal{P}= [-0.22, \, 1]$. Now, as indicated in Section~\ref{sec:implementation}, we implement~\eqref{eq:unbalanced_disk} as a \lstinline{nlss} object and execute the LPV embedding with the function \lstinline{nlss2lpvss} specifying an analytical integration of \eqref{eq:simplified_factorized_nl}. By specifying the input argument \mbox{\lstinline{'factor'}}, \lstinline{nlss2lpvss} produces the same solution as in~\eqref{eq:unbalanced_LPV_manual}. With the input argument \mbox{\lstinline{'element'}}, the following equivalent LPV model is obtained
\begin{subequations} \label{eq:unbalanced_LPV_auto}
    \begin{equation}
        \begin{aligned}
            \dot{x}(t) & = \begin{bmatrix}
                               0 & 1 \\ p(t) & -\frac{1}{\tau}
                           \end{bmatrix}x(t) + \begin{bmatrix}
                                                   0 \\ \frac{K_\mathrm{m}}{\tau}
                                               \end{bmatrix} u(t); \\
            y(t)       & = \begin{bmatrix}
                               0 & 1
                           \end{bmatrix} x(t),
        \end{aligned}
    \end{equation}
    with
    \begin{align}
        p(t) & = \frac{M g l}{J} \operatorname{sinc}(x_1(t)),
    \end{align}
\end{subequations}
resulting in a $\mathcal{P}=[-28.45, \, 130.96]$.
To show what happens if symbolic integration is substituted with numerical integration of~\eqref{eq:simplified_factorized_nl}, an LPV model $\Pi_\mathrm{LPV}^\mathrm{num}$ is produced. Note that, due to symbolic integration in Matlab, the $\operatorname{sinc}$ function in the scheduling map is expressed as $\frac{\sin(x_1)}{x_1}$, which is theoretically well-defined in the limit, but leads to computational issues when directly evaluated at $x_1 = 0$. This issue can be solved in the software by redefining the produced scheduling map as a piecewise function, where the output at $x_1 = 0$ is explicitly defined using limit.

As the proposed LPV embedding approach is exact, the solutions of the nonlinear system are expected to exactly match those of the LPV models produced.
To verify this, we compare the simulation of the nonlinear system $\Pi_\mathrm{NL}$ in~\eqref{eq:unbalanced_disk} with the self-scheduled simulation of the LPV models $\Pi_\mathrm{LPV}^\mathrm{ana}$ and $\Pi_\mathrm{LPV}^\mathrm{num}$ produced by \lstinline{nlss2lpvss}. We use the \textsc{Matlab} in-built variable-step solver \lstinline{ode45} with the default parameters and step size, and simulate for 15 seconds.
\begin{figure}[t]
    \centering
    \includegraphics[width=\columnwidth]{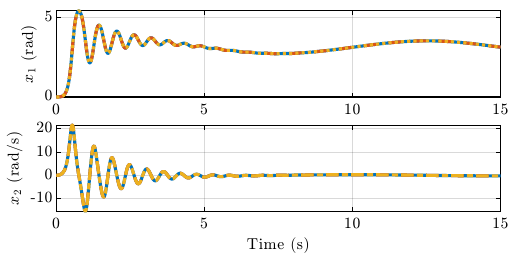}
    \caption{Simulation of the nonlinear unbalanced disk $\Pi_\mathrm{NL}$ (\crule{0, 0.447, 0.7410}{8pt}{1.2pt}), self-scheduled simulation of the LPV models $\Pi_\mathrm{LPV}^\mathrm{ana}$ (\crule{0.85, 0.325, 0.098}{4pt}{1.2pt}\,\crule{0.85, 0.325, 0.098}{4pt}{1.2pt}) and  $\Pi_\mathrm{LPV}^\mathrm{num}$ (\crule{0.929, 0.694, 0.125}{3pt}{1.2pt}\,\crule{0.929, 0.694, 0.125}{1pt}{1.2pt}\,\crule{0.929, 0.694, 0.125}{3pt}{1.2pt}).}
    \label{fig:sim_unbaldisk_LPVembedding}
\end{figure}
The system is initialized with the mass at the upwards position with zero velocity, i.e. $x(0) = \operatorname{col}(0,0)$, and we consider $u(t) = 2 \sin(0.2 \pi t)$ as the input to the system. The results in Fig.~\ref{fig:sim_unbaldisk_LPVembedding} show that the solution of the nonlinear system is equivalent to the solutions of the LPV models.
The \emph{root-mean-square error} (RMSE) between the state responses of the nonlinear simulation and the LPV models are reported in Table.~\ref{tab:RMSE_lpvembeddings}, indicating a negligible error due to numerical precision.
\begin{table}[t]
    \centering
    \caption{RMSE between the simulation of the nonlinear model $\Pi_\mathrm{NL}$ and the self-scheduled simulation of the LPV models $\Pi_\mathrm{LPV}^\mathrm{ana}$ and $\Pi_\mathrm{LPV}^\mathrm{num}$.}
    \begin{tabular}{l l l}
        \hline
                                                              & \textbf{State} ($x_1$) & \textbf{State} ($x_2$) \\
        \hline
        \vspace{1mm} RMSE for $\Pi_\mathrm{LPV}^\mathrm{ana}$ & \num{3.18e-13}         & \num{3.67e-12}         \\
        RMSE for $\Pi_\mathrm{LPV}^\mathrm{num}$              & \num{5.82e-14}         & \num{6.67e-13}         \\[0.5mm]
        \hline
    \end{tabular}
    \label{tab:RMSE_lpvembeddings}
\end{table}

\begin{figure}[t]
    \centering
    \includegraphics[width=0.85\columnwidth]{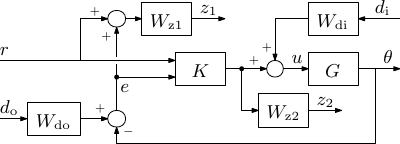}
    \caption{Generalized plant structure for the controller synthesis of the unbalanced disk.}
    \label{fig:genSS}
\end{figure}

\begin{figure}[b]
    \centering
    \includegraphics[width=0.7\columnwidth]{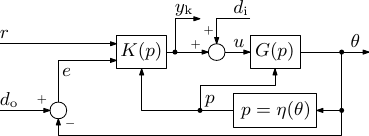}
    \caption{Schematic representation of the self-scheduled closed-loop simulation.}
    \label{fig:clic_genss}
\end{figure}

\begin{figure}[t]
        \centering
        \includegraphics[width=\columnwidth]{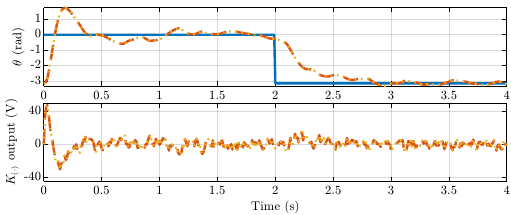}
        \caption{At the top, the angle of the unbalanced disk in closed-loop with the controllers $K^\mathrm{ana}$ (\crule{0.85, 0.325, 0.098}{4pt}{1.2pt}\,\crule{0.85, 0.325, 0.098}{4pt}{1.2pt}) and $K^\mathrm{num}$ (\crule{0.929, 0.694, 0.125}{3pt}{1.2pt}\,\crule{0.929, 0.694, 0.125}{1pt}{1.2pt}\,\crule{0.929, 0.694, 0.125}{3pt}{1.2pt}) under the reference (\crule{0, 0.447, 0.7410}{8pt}{1.2pt}). At the bottom, the inputs to the plant generated by the controllers.}
        \label{fig:clic_sim}
    \end{figure}

Lastly, we test the obtained LPV models for LPV controller synthesis. For this, we construct the generalized plant structure 
as depicted in Fig.~\ref{fig:genSS},
using the \textsc{LPVcore} extension of the \textsc{Matlab} function \lstinline{connect}. In the generalized plant, $G$ is an LPV representation of the unbalanced disk, $K$ is the to-be-synthesized LPV controller, $r$, $d_\mathrm{i}$ and $d_\mathrm{o}$ are disturbance channels and $z_1$ and $z_2$ the generalized performance channels. Specifically, $r$ is the reference, $d_\mathrm{i}$ is an input load and $d_\mathrm{o}$ is an output disturbance. The controller $K$ is designed in a two-degrees of freedom configuration, corresponding to a joint design of a feedforward and a feedback controller. This can be seen from the reference trajectory and tracking error being two separate inputs to the controller. The weighting filters are chosen as
    \begin{align}
         & W_{\mathrm{z1}}(s)  = \frac{0.5012s + 3.0071}{s + 0.0301}; &  & W_\mathrm{z2} (s)  = \frac{10s + 400}{s + 4000}, \nonumber \\
         & W_\mathrm{di}       = 0.5;                                 &  & W_\mathrm{do}      = 0.1,
    \end{align}
    where $W_{\mathrm{z1}}$ has low-pass characteristics to approximate integral action and ensure good tracking performance at low frequencies, $W_{\mathrm{z2}}$ has high-pass characteristics to enforce roll-off at high frequencies, and each filter $W_\mathrm{d\ast}$ considers an upper bound on the expected magnitude of the input disturbances.
    Next, we invoke the \textsc{LPVcore} \lstinline{lpvsyn} function to synthesize the LPV controllers $K^\mathrm{ana}$ and $K^\mathrm{num}$ using an $\mathcal{L}_2$-gain optimal polytopic LPV controller synthesis approach \cite{APKARIAN19951251}. Each controller is synthesized by using either $\Pi_\mathrm{LPV}^\mathrm{ana}$ with $\mathcal{P}=[-0.22, \, 1]$ or $\Pi_\mathrm{LPV}^\mathrm{num}$ with $\mathcal{P}=[-28.45, \, 130.96]$ as $G$ in the generalized plant. Both controllers result in an $\mathcal{L}_2$-gain of $\gamma = 0.954$. Then, we test the resulting controllers with a closed-loop self-scheduled simulation, as illustrated in Fig.~\ref{fig:clic_genss}.
Now, we use the \textsc{Matlab} variable-step solver \lstinline{ode15s} with the default settings and step size as the closed-loop interconnection becomes a stiff system, and simulate for 4 seconds. The system is initialized with the mass at the downwards position with zero velocity, i.e. $x(0) = \operatorname{col}(\pi, 0)$, while the controller states are initialized at zero. The input disturbances are generated as discrete-time signals with a sampling time of $T_s = 0.01 \ (s)$, with $d_i(t) \sim  \mathcal{U}(-0.5, 0.5)$ and $d_o(t)\sim  \mathcal{U}(-0.1, 0.1)$ where $\mathcal{U}$ denotes a uniform distribution, and are applied in a zeroth-order hold setting. In Fig.~\ref{fig:clic_sim}, the output of the unbalanced disk is shown together with the controller output for a reference signal that induces a swing-up motion of the mass at time zero, and recovers the downwards position afterwards. The difference between $K^\mathrm{ana}$ and $K^\mathrm{num}$ is negligible and both achieve the desired reference tracking and disturbance rejection objectives.
    %
    %

    %

    \subsection{3DOF control moment gyroscope}
\begin{figure}[b]
        \centering
        \includegraphics[width=0.45\columnwidth]{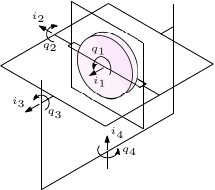}
        \caption{Schematics of the Quanser 3DOF control moment gyroscope, where the angular positions are indicated by $q_i$ and the motor inputs by $i_i$.}
        \label{fig:geom_gyro}
    \end{figure}

    The 3DOF CMG consists of three actuated gimbals and a flydisk mounted on the inner gimbal, as illustrated in Fig.~\ref{fig:geom_gyro}.
    The nonlinear dynamics that describe the CMG are presented in~\citep{5841d138cdd94f17aaa413781677ac2a}, resulting in the differential equation
    \begin{equation}
        M(q(t))\ddot{q}(t) + \left( C\left(q(t), \dot{q}(t) \right) + F_\mathrm{v} \right) \dot{q}(t) = K_\mathrm{m} u(t),
    \end{equation}
    where $t \in \R$ is time, $q(t) \in \R^4$ are the angular positions of the gimbals, $i(t) \in \R^4$ are the input currents to the motors, $M$ is the inertia matrix, $C$ is the Coriolis matrix, $F_\mathrm{v}$ is the viscous friction matrix, and $K_\mathrm{m}$ is the motor gain matrix. For this example, we consider that the third gimbal is locked, i.e. $q_3, \ \dot{q}_3 \equiv 0$ 
    and only the first two gimbals are actuated, i.e. $i_3, \ i_4 \equiv 0$. 
    As outputs of the system, we consider the angular position of the unlocked gimbals, $y = \operatorname{col}\left(q_1, q_2, q_4 \right)$. Then, the dynamics of the CMG system can be represented by a continuous-time nonlinear model of the form of~\eqref{eq:nl_dyn}, where $x(t) = \operatorname{col}\left(q_1, q_2, q_4, \dot{q}_1, \dot{q}_2, \dot{q}_4 \right)(t) \in \R^6$, $u(t) = \operatorname{col}\left(i_1, i_2 \right)(t) \in \R^2$, and $y(t) = \operatorname{col}\left(q_1, q_2, q_4 \right)(t) \in \R^3$.

\begin{figure}[t]
    \centering
    \includegraphics[width=\columnwidth]{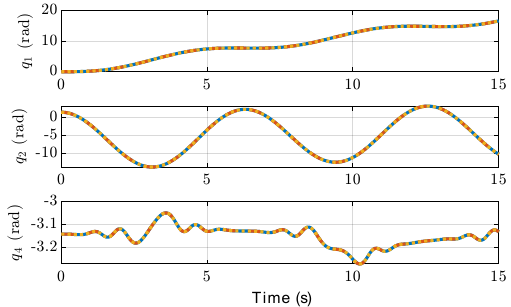}
    \caption{Simulation of the nonlinear CMG (\crule{0, 0.447, 0.7410}{8pt}{1.2pt}), self-scheduled simulation of the LPV model $\Omega_\mathrm{LPV}^\mathrm{ana}$ (\crule{0.85, 0.325, 0.098}{4pt}{1.2pt}\,\crule{0.85, 0.325, 0.098}{4pt}{1.2pt}) and the LPV model $\Omega_\mathrm{LPV}^\mathrm{num}$ (\crule{0.929, 0.694, 0.125}{3pt}{1.2pt}\,\crule{0.929, 0.694, 0.125}{1pt}{1.2pt}\,\crule{0.929, 0.694, 0.125}{3pt}{1.2pt}).}
    \label{fig:sim_gyro_LPVembedding}
\end{figure}

    Now, two LPV embeddings $\Omega_\mathrm{LPV}^\mathrm{ana}$ and $\Omega_\mathrm{LPV}^\mathrm{num}$ of the CMG model are obtained using \lstinline{nlss2lpvss} with the respective options \lstinline{'analytical'} and \lstinline{'numerical'}, and both using \lstinline{'factor'}. In this case, while $\Omega_\mathrm{LPV}^\mathrm{ana}$ has $n_\mathrm{p} = 15$ scheduling variables, $\Omega_\mathrm{LPV}^\mathrm{num}$ contains only $n_\mathrm{p} = 13$. Note that this difference in scheduling dimensions is caused by the extraction process of the nonlinear elements from the matrix functions of the system, but both LPV models should be equivalent. Moreover, if required for any downstream tasks (such as controller synthesis), the scheduling dimension can be reduced using the LPV-SS scheduling dimension reduction methods present in \textsc{LPVcore}. To verify that the produced LPV models are equivalent to the original CMG model, we compare the simulation of the CMG model with the self-scheduled simulation of $\Omega_\mathrm{LPV}^\mathrm{ana}$ and $\Omega_\mathrm{LPV}^\mathrm{num}$. The simulation is run for 15 seconds using the \textsc{Matlab} variable step solver \mbox{\lstinline{ode15s}} with the default settings and step size. The gimbals are initialized at the angular position of $(q_1(0), q_2(0), q_4(0)) = (0, \pi/2, -\pi)$ with zero angular velocity, and we consider $i_1(t) = 0.1 \sin(t)$ and $i_2(t) = -0.3 \cos(t)$ as inputs to the system. In Fig.~\ref{fig:sim_gyro_LPVembedding}, the simulation results show that the output trajectories are equivalent. Moreover, we compute the RMSE error between the simulation outputs of the original CMG model and the LPV models and it resulted in less than \num{1e-12} in all cases.

%% file: chapters/conclusion.tex
\section{Conclusion\label{sec:conclusion}}
This paper has introduced an automated LPV model conversion approach for nonlinear dynamical systems
together with discussing its implementation in the \textsc{LPVcore} toolbox and demonstrating its  capabilities via simulation and controller design examples. Compared to other LPV model conversion approaches, the key difference is that the proposed method achieves an automated LPV embedding without any approximation or need for exploring combinatorial options in the substitution or conversion of nonlinear terms, providing a readily usable solution for practical LPV modeling problems. The \textsc{LPVcore} toolbox is freely available on \url{www.lpvcore.net}, and the examples used in Section~\ref{sec:examples} can be found in \url{https://gitlab.com/Javi-Olucha/lpvs25-code-repo}.